\documentclass[numbers]{article}
\usepackage{amsfonts,amsmath,amssymb}
\usepackage[mathscr]{euscript}
\usepackage[numbers]{natbib}
\usepackage{hyphenat}
\usepackage[english]{babel}

\usepackage[latin1]{inputenc}
\usepackage[T1]{fontenc}
\usepackage{lmodern}

\newcommand{\dcauthorsurname}{Trevi\~no Aguilar} 
\newcommand{\dcauthorname}{Erick} 

\newcommand{\dctitle}{T-Systems and the lower Snell envelope}

\newcommand{\dckeyena}{Lower Snell Envelope}
\newcommand{\dckeyenb}{Regularization of stochastic processes}
\newcommand{\dckeyenc}{Robust optimal stopping}
\newcommand{\dckeyend}{Stability under pasting}

\usepackage{ifpdf}

\ifpdf
\usepackage[%
	pdftitle={\dctitle},
	pdfauthor={\dcauthorsurname\ \dcauthorname},
	pdfkeywords={\dckeyena, \dckeyenb, \dckeyenc, \dckeyend},
	pdfpagemode=UseOutlines,
  colorlinks=true,					% bitte nicht ändern!
	linkcolor=black,					% bitte nicht ändern!
	filecolor=black,					% bitte nicht ändern!
	urlcolor=black,						% bitte nicht ändern!
	citecolor=black,					% bitte nicht ändern!
	pdftex=true,              % bitte nicht ändern!
	plainpages=false,         % bitte nicht ändern!
	hypertexnames=false,      % bitte nicht ändern!
	pdfpagelabels=true,       % bitte nicht ändern!
	hyperindex=true]{hyperref}% bitte nicht ändern!
\else
\fi

\newcommand{\h}[1][H]{ \ensuremath{\{#1_t\}_{0 \leq t \leq T}}}	 	% A randomized test
\newcommand{\ls}[1][t]{ \ensuremath{U^{\downarrow}_{#1}}} 	%Lower Snell envelope
\newcommand{\lsv}[1][t]{ \ensuremath{Z^{\downarrow}_{#1}}} 	%variable Lower Snell envelope

\newcommand{\ftp}{ \ensuremath{\mt}} 			% Familia de tiempos de paro 
\newcommand{\tp}{ \ensuremath{\theta}}	 	% Tiempo de paro tipico
\newcommand{\nc}[3][Corollary]{
\newtheorem{#2}[d1]{#1}
\begin{#2}  \label{#2}
#3
\end{#2}
}

\newcommand{\fnd}[3][Definition]{
\newtheorem{d1}{#1}[section] 
\begin{d1}  \label{#2}
#3
\end{d1}
}
\newcommand{\nd}[3][Definition]{
\newtheorem{#2}[d1]{#1}
\begin{#2}  \label{#2}
#3
\end{#2}
}

\newcommand{\eq}[2]{\begin{equation} \label{#1} 
#2 \end{equation}}
\newcommand{\nl}[3][Lemma]{
\newtheorem{#2}[d1]{#1}
\begin{#2}  \label{#2}
#3
\end{#2}
}
\newcommand{\np}[3][Proposition]{
\newtheorem{#2}[d1]{#1}
\begin{#2}  \label{#2}
#3
\end{#2}
}

\newcommand{\nt}[3][Theorem]{
\newtheorem{#2}[d1]{#1}
\begin{#2}  \label{#2}
#3
\end{#2}
}

\newcommand{\nr}[3][Remark]{
\newtheorem{#2}[d1]{#1}
\begin{#2}  \label{#2}
#3
\end{#2}
}
\newcommand{\fincor}{\ensuremath{\square}}  %fin corollary
\newcommand{\finlemma}{\ensuremath{\square}}			%fin lema	
\newcommand{\fintheo}{\ensuremath{\square}}	%fin teoremas
\newcommand{\ps}{\ensuremath{(\Omega, \mathcal{F}, \mathbb{F}=\{\mathcal{F}_t\}_{t \in [0,T]}, R)}} % base estocastica
\newcommand{\li}[1][i]{\ensuremath{ \lim_{#1 \to \infty}}} 	%limite hasta el infinito
\newcommand{\lis}[1][i]{\ensuremath{ \mathrm{lim \thinspace sup}_{#1 \to \infty}}} 	%lim sup limite superior hasta el infinito
\newcommand{\lii}[1][i]{\ensuremath{ \mathrm{lim \thinspace inf}_{#1 \to \infty}}} 	%lim inf limite inferior hasta el infinito
\newcommand{\essinf}{ \ensuremath{\mathrm{ess \thinspace inf}}} 	%essential infimum
\newcommand{\esssup}{ \ensuremath{\mathrm{ess \thinspace sup}}} 	%essential supremum

\newcommand{\indf}[1][A]{\ensuremath{ 1_{\{#1\}}}} 	%indikator function

\newcommand{\seq}[1][P]{ \ensuremath{ \{{#1}_i\}_{i=1}^{\infty}  } } %sucesion indice abajo
\newcommand{\seqa}[1][P]{ \ensuremath{ \{{#1}^i\}_{i=1}^{\infty}  } } %sucesion indice arriba

\newcommand{\rcll}{ \mbox{c\`adl\`ag}} 	%procesos cadlag
\newcommand{\envelop}{\mbox{envelope}}

\newcommand{\mbf}{\ensuremath{\mathbb{F}}}

\newcommand{\mbn}{\ensuremath{\mathbb{N}}}

\newcommand{\mf}{\ensuremath{\mathcal{F}}}

\newcommand{\mq}{\ensuremath{\mathcal{Q}}}

\newcommand{\mt}{\ensuremath{\mathcal{T}}}

\normalsize

\begin{document}
%------------------------------------------------------------------------------------------------------------------------
%*********************************************************************************************************************************************
\title{$\ftp$-Systems and the lower Snell envelope}
\maketitle
\begin{center}
\author{{\bf Trevi\~no-Aguilar Erick\footnote{Centro de Investigaci\'on en Matem\'aticas A.C., Guanajuato M\'exico. email trevino@cimat.mx}}}
\end{center}
%------------------------------------------------------------------------------------------------------------------------
\begin{abstract}
The dynamical analysis of American options has motivated the development of robust versions of the classical Snell envelopes. The cost of superhedging an American option is characterized by the upper Snell envelope.  The infimum of the arbitrage free prices  is characterized by the lower Snell envelope. In this paper we focus on the lower Snell envelope.  We construct a regular version of this stochastic process. To this end, we apply results due to Dellacherie and Lenglart on regularization of stochastic processes and $\ftp$-Systems.
\end{abstract}
%------------------------------------------------------------------------------------------------------------------------
\paragraph{Keyword:} Lower Snell Envelope, Regularization of stochastic processes, Robust optimal stopping, Stability under pasting.

%*******************************************************************************************************************
\section{Introduction}
%*******************************************************************************************************************
American options allow for the possibility of early liquidation. From the point of view of the buyer derives the problem of optimal exercise. It is well understood in the context of complete financial markets, that is, when the market admits a unique martingale measure $P^*$ for the price process; see Bensoussan\cite{Bensoussanpricing} and Karatzas\cite{Karatzaspricingao}. The key to the solution is provided by the construction of the so called Snell envelope: The smallest $P^*$-supermartingale dominating the payoff process of the American option. The option can be  optimally exercised when the payoff process touches the Snell envelope.  From the point of view of the seller, the Snell envelope characterizes the hedging strategy through the martingale part of the Doob-Meyer decomposition and the corresponding stochastic representation. In the context of incomplete markets, the analysis is substantially more complicated  since there is a family of martingale measures. The analysis of American options in incomplete markets has motivated the development of robust versions of the Snell envelope.  The  superhedging cost of American options is  characterized by the upper Snell envelope, due to the Optional Decomposition Theorem; see F\"ollmer and Kramkov \cite{Follmerkramkov}.	The infimum of the arbitrage free prices  is characterized by the lower Snell envelope by F\"ollmer and Schied\cite{Follmerschied}, Theorem 6.33,  in a general discrete-time model,  and by Karatzas and Kou\cite{Karatzaskou}, Theorem 5.13, in a continuous-time model driven by Brownian motion.\\
 
The lower Snell envelope  appears in other contexts such as the optimal exercise of American options. In this context, the preferences of the buyer are explicitly taken into account and represented through a robust utility functional 
$$
\psi(\cdot):=\inf_{Q \in \mq}E_Q[u(\cdot)],
$$
with $\mq$ a convex class of equivalent probability measures and $u$  a concave utility function. Thus, preferences on the face of risk are  quantified as clarified by the robust extension of the classical Neumann-Morgenstern Theory\cite{Neumannmorgenstern} due to Gilboa and Schmeidler\cite{Gilboaschmeidler}. An American option with payoff process $H:=\h$  has the maximal robust utility
$$
\sup_{\tp \in \ftp} \psi(H_{\tp}),
$$
where the supremum is taken over the family of stopping times of the trading period. This approach to optimal exercise, and the role of the lower Snell envelope,  is discussed by F\"ollmer and Schied\cite{Follmerschied}  in discrete time for the special case where $\mq$ is a \textit{stable} family of equivalent probability measures. The axiomatic framework of this special class of preferences, and the corresponding robust representation for the preference order,  is due to Epstein and Schneider\cite{Epsteinschneider}.\\  

Other motivation for the lower Snell envelope arise from a game theoretic point of view; see e.g., Zamfirescu\cite{Zamfirescu}.  Riedel\cite{Riedel} studies a problem of optimally stopping a process in discrete time when model uncertainty is explicitly taken into account.\\

In this paper we focus in the lower Snell envelope.  Our main goal is to construct a regular version of this process. More precisely, we show how to apply the theory of regularization of stochastic processes and $\ftp$-Systems, due to Dellacherie and Lenglart\cite{Dellacherielenglartrecollement}, in order to obtain a $\rcll$-version of the lower Snell envelope.\\

The rest of the paper is organized as follows. In Section \ref{labelsectionlowersnellenvelope} we formally introduce the lower Snell envelope of a stochastic process $H$ given that a stable family of equivalent probability measures $\mq$ is fixed. We then present the main result of the paper, Theorem \ref{teoconstructionse}. The proof will need some preparation, this is distributed in the remaining sections.  In subsection \ref{labelsectionclassicalstopping}, we recollect a general result of optimal stopping and the classical Snell envelope. In subsection \ref{labelsectionstabilityunderpasting}, we recollect general results about the property of stability. In Section \ref{labelsectionoptimalstoppingtimes}, we solve a robust stopping problem involving the class of probability measures $\mq$; see Proposition \ref{lspls}. In Section \ref{labelsectiontsystems} we introduce the concept of $\ftp$-Systems and recollect the results that we are going to apply. In Section \ref{labelsectionlaprueba} we conclude the proof of Theorem \ref{teoconstructionse}.
%*******************************************************************************************************************
\section{The lower Snell envelope} \label{labelsectionlowersnellenvelope}
%*******************************************************************************************************************
We start with some notation. We fix a stochastic base with finite horizon
$$\ps.$$
The probability measure $R$ is a reference measure, and we assume it is $0-1$ in $\mf_0$. We assume that  the filtration $\mbf$ satisfies the usual assumptions of right continuity and completeness. By $\mt$  we denote the class of $\mbf$-stopping times  with values in $[0,T]$. For a stopping time $\tau \in \mt$ we define $\mt[\tau,T]:= \{\tp \in \mt \mid R(\tp \geq \tau)=1\}$.\\

We fix a family of equivalent probability measures $\mq$ which is stable in the sense of the following definition.
\fnd[Stability under pasting]{defstability}
{
Let $\tau  \in \ftp$ be a stopping time  and $Q_{1}$ and $Q_{2}$ be probability measures equivalent to $R$. 
The probability measure defined through
$$Q_3(A):=E_{Q_1}[Q_{2}[A \mid \mf_{\tau}]],A \in \mf_T$$
is called {\rm the  pasting} of $Q_1$ and $Q_2$ in $\tau$.\\

The family of  probability measures   $\mq$ is  {\rm stable under pasting} or simply {\rm stable} if every $Q \in \mq$ is equivalent to $R$, and if for each $Q_{1}$ and $Q_{2}$ in $\mq$ and any stopping time $\tau \in \ftp$,  the pasting of $Q_1$ and $Q_2$ in $\tau$ is an element of $\mq$.
}
Notice that stability is only formulated for  families whose elements are  equivalent to the reference probability measure $R$.  We taked Definition \ref{defstability} from F\"ollmer and Schied\cite{Follmerschied}. It is related to the concepts of fork-convexity and  m-stability; see e.g., Delbaen\cite{Delbaenastablesets}.  F\"ollmer and Schied\cite{Follmerschied}   clarify the role of stability of the family of equivalent martingale measures for the analysis of the upper and  lower prices $\pi_{\sup}(\cdot)$ and $\pi_{\inf}(\cdot)$  of American options  in discrete time.  Another important application of the stability concept appears in the problem of representing   dynamically consistent   risk measures; see e.g.,  F\"ollmer and Penner\cite{Follmerpenner} for details and references.\\

We precise the payoff process $H$ of the introduction: It is a $\rcll$ positive $\mbf$-adapted process. We assume that the process $H$ is of $class(D)$ with respect to each  $Q \in \mq$, thus 
$$
\lim_{x  \to \infty} \sup_{\tp \in \ftp}  E_Q[H_{\tp}; H_{\tp} \geq x]=0.
$$
In particular
\eq{eqintlsepindiidual}
{
\sup_{\tp \in \ftp} E_Q[H_{\tp}] < \infty.
}
Moreover, the process $H$ is regular in the sense of the following definition.  The concept is motivated by Definition 2.11 and Remark 2.42 of  El Karoui\cite{{Karouiaspectsprobabilistescs}}.
\nd{defuscinex}
{
The stochastic  process  $H$ is said to be {\rm upper semicontinuous in expectation from the left with respect to the probability measure} $Q$ if for any increasing  sequence of 
stopping times $\seq[\tau]$ converging to $\tau$, we have
\begin{equation}
\label{lesce}
\limsup_{i \rightarrow  \infty} E_Q[H_{\tau_i}] \leq E_Q[H_{\tau}].
\end{equation}
}
For $\tau$ a stopping time we define
\begin{align}
Z_{\tau}^Q&:= \esssup_{\tp \in \ftp[\tau,T]} E_Q[H_{\tp} \mid \mf_{\tau}], \notag \\
\lsv[\tau]&:= \essinf_{Q \in \mq} Z^Q_{\tau}= \essinf_{Q \in \mq} \esssup_{\tp \in \ftp[\tau,T]} E_Q[H_{\tp} \mid \mf_{\tau}].
\label{eqlsv}
\end{align}
\nd{labellowersenellenvelop}
{
The {\rm lower Snell envelope} of $H$ with respect to the stable class $\mq$  is the stochastic process defined by 
\eq{eqdeflsv}
{
\lsv[]:=\{\lsv[t] \}_{0\leq t\leq T}.
}
}

The main result of the paper is the following 
\nt{teoconstructionse} 
{
There exists an optional right-continuous stochastic process $\{\ls\}_{0\leq t \leq T}$ such that for any stopping time  $\tp \in \ftp$ 
\[
\ls[\tp]=\lsv[\tp], \thinspace R-a.s.
\]
In particular, $\ls[]$ is a modification of the lower Snell envelope $\lsv[]$.
}
As guideline for notation, we emphasize that $\lsv[\tp]$ should be interpreted as a random variable associated to the stopping time $\tp$, while $\ls[\tp]$ is a stochastic process sampled in the stopping time $\tp$. Note also that the stochastic process \eqref{eqdeflsv} is adapted, but we do not have any property of regularity not of  measurability. In particular, a construction like:
\[
\inf\{t \geq 0 \mid \lsv \geq H_t\},
\]
does not necessarily produce a stopping time in a general model.\\

Let us comment on the strategy we follow to prove Theorem \ref{teoconstructionse}.  For the first part of the proof  we fix a stopping time $\rho \in \ftp$. In Proposition \ref{lspls} we construct a stopping time $\tau^{\downarrow}_{\rho}$ such that 
\[
\lsv[\rho]= \essinf_{Q \in \mq}E_{Q}[H_{\tau^{\downarrow}_{\rho}}\mid \mf_{\rho}].
\]
This allow us to conclude that 
\eq{labelequationmm}
{
\lsv[\rho]= \esssup_{\tp \in \ftp[\rho,T]}\essinf_{Q \in \mq}E_{Q}[H_{\tp}\mid \mf_{\rho}];
}
see Corollary \ref{lmmls}.  In the second part of the proof, we prove that the family of random variables 
\[
\{\lsv[\tp]\}_{\tp \in \ftp},
\]
is a $\ftp$-System; see Definition \ref{defdellacherie} and Lemma \ref{lemmaftpsystem}. We then use the expression \eqref{labelequationmm} to prove that this $\ftp$-System is right-continuous; see Definition \ref{defuppsemcon} and Lemma \ref{lemmaftpsystemrightcontinuous}.  We conclude the proof with the Corollary \ref{corrightcontinuity}.
%************************************************************************************************************************
\subsection{The classical non-robust stopping problem}\label{labelsectionclassicalstopping}
%************************************************************************************************************************
The  solution of the classical non-robust stopping problem through the Snell envelope is the content of the next theorem.  It will play a key role in the solution of the robust case.   We fix a probability measure $Q \in \mq$.  Note that in this theorem we consider starting points other than $t=0$.
\nt{theosekaroui}
{
\begin{enumerate}
\item There exists a $\rcll$ supermartingale denoted $U^Q(H)$, or simply $U^Q$,   such that
$$U^Q_{\tau}= \esssup_{\tp \in \ftp[\tau,T]} E_Q[H_{\tp} \mid \mf_{\tau}],\quad Q-a.s.,$$
for any stopping time $\tau \in \ftp$. $U^Q$ is the minimal $\rcll$ supermartingale that dominates $H$.  $U^Q$ is of $class(D)$ due to the fact that $H$ is of $class(D)$.
\item Let $\rho \in \ftp$. A stopping time $\tau^* \in \ftp[\rho,T]$ is optimal in the sense that 
$$U^Q_{\rho} = E_Q[H_{\tau^*} \mid \mf_{\rho}],\quad Q-a.s.,$$
if and only if
\begin{enumerate}
\item The process $\{U^Q_{s \wedge \tau^*}\}_{{\rho} \leq s \leq T}$ is a martingale, and
\item $H_{\tau^*} = U^Q_{\tau^*},\quad Q-a.s.,$
\end{enumerate}
\item Optimal stopping times exist and the minimal one is given by
\eq{eqtautkarouidinamico}
{
\tau^Q_{\rho}	:=\inf \{s \geq \rho \mid H_s \geq U^Q_s\}.
}
\end{enumerate}
}
Proof.
See Theorems 2.28, 2.31, 2.39  and 2.41  in  El Karoui\cite{Karouiaspectsprobabilistescs}.\fintheo\\

\nd{dse}
{
The stochastic process $U^Q$ constructed in  Theorem \ref{theosekaroui} is called the {\rm Snell $\envelop$} of $H$ with respect to $Q$.
}
%************************************************************************************************************************
\subsection{Stability under pasting} \label{labelsectionstabilityunderpasting}
%************************************************************************************************************************
The stability of the family $\mq$ is crucial for the next  lemmas to hold true.   They are  versions in  continuous time of the analysis of F\"ollmer and Schied\cite{Follmerschied}, Section 6.5. The first lemma will be necessary in the construction of optimal robust stopping times. The second and third lemmas  will be used in the construction of a right-continuous version of the lower Snell envelope; see Lemma \ref{lemmaftpsystemrightcontinuous}.

\nl{lccep}
{
Let $Q_3$ be the pasting of $Q_1$ and $Q_2$ in $\sigma$.  Let $Y$ be a positive random variable $\mf_T$-measurable and $Q_i$-integrable for $i=1,2,3$.
Then, for any stopping time $\tau \in \ftp$ we have 
\[
E_{Q_3}[Y\mid \mf_{\tau }]=E_{Q_{1}}[E_{Q_{2}}[Y\mid \mf_{\sigma \vee \tau }]\mid \mf_{\tau }].\finlemma
\]
}

For the second lemma it is convenient to introduce the notation:
\eq{eqstableinitcond}
{
\mq(Q_0,\tau):=\{Q \in \mq \mid Q=Q_0 \mbox{ in } \mf_{\tau}\}, \textrm{ for } Q_0 \in \mq \textrm{ and } \tau \in \ftp.
}
\nl{lfplll}
{
Let $Q_0 \in \mq$ be  arbitrary but  fixed. Then, for  stopping times $\sigma, \tau, \tp \in \ftp$ with $\sigma \leq \tau \leq \tp$ we have
\eq{loque}
{
E_{Q_0}[\essinf_{Q \in \mq} E_Q[H_{\tp}\mid \mf_{\tau}]\mid \mf_{\sigma}] =\essinf_{Q \in \mq(Q_0,{\tau})} E_{Q}[H_{\tp}\mid \mf_{\sigma}].\finlemma
}
}

\nl{lemabackwardssubmarrobust}
{Let $Y$ be a positive random variable $\mf_T$-measurable such that
\[
E_Q[Y] < \infty,
\]
for each $Q \in \mq$. Let $\rho \in \ftp$. Let  $\seq[\rho] \subset \ftp$ be a decreasing sequence of stopping times converging to $\rho$. Then, the sequence of random variables $\seq[Y]$ defined by
\[
Y_i:= \essinf_{Q \in \mq} E_Q[Y \mid \mf_{\rho_i}],
\] 
is a backward $\mq$-submartingale in the following sense: For any $Q \in \mq$ and $i \in \mbn$
\eq{eqlemabacckweraddsubrocbuuno}
{
E_Q[Y_{i} \mid \mf_{\rho_{i+1}}] \geq Y_{i+1}, \quad Q-a.s.
}
Moreover, 
\eq{eqlimabackubrocklimi}
{
\li Y_i
}
exists $R$-a.s. and in $L^1(Q)$ for any $Q \in \mq$.
}
%************************************************************************************************************************
\section{Optimal robust stopping times}\label{labelsectionoptimalstoppingtimes}
%************************************************************************************************************************
Recall that $H$ is a  process of $class(D)$ and is upper semicontinuous in expectation from the left with respect to each $Q \in \mq$. The stopping time $\tau^Q_{\rho}$ was defined in \eqref{eqtautkarouidinamico}.\\  

Let us comment Proposition \ref{lspls} below.  In the definition \eqref{ligual} of the random variable $\tau^{\downarrow}_{\rho}$, we need to verify that this random variable is in fact a stopping time. This is non trivial and is the first part of the proposition.  The second part of the proposition extends a result of Karatzas and Kou\cite{Karatzaskou}, Formula (5.33). The extension consist in the facts that we consider a genreal model and we do not use an apriori regularity property of the lower Snell envelope. Instead, we use the stability of the family $\mq$.
\np{lspls}
{
Let $\rho \in \ftp$ be fixed. The random time
\eq{ligual}
{
\tau^{\downarrow}_{\rho}:=\essinf_{Q \in \mq} \tau^Q_{\rho},
}
is a {\rm stopping time}. Moreover, it is optimal in the following sense
\eq{lespmm}
{
\lsv[\rho]= \essinf_{Q \in \mq}E_{Q}[H_{\tau^{\downarrow}_{\rho}}\mid \mf_{\rho}].
}
In particular for  $\tau^{\downarrow}_{0}$: 
\eq{tmaxroutst}
{
\inf_{Q \in \mq} E_{Q}[H_{\tau^{\downarrow}_{0}}]=
\sup_{\tp \in \ftp}\inf_{Q \in \mq} E_{Q}[H_{\tp}].
}
}
Proof. The optimality of $\tau^Q_{\rho}$ with respect to $Q$  follows from Theorem \ref{theosekaroui}.
\begin{enumerate}
\item First  we prove that \eqref{ligual}  indeed defines a stopping time.  To this end, we show that the family $\{\tau^Q_{\rho}\}_{Q \in \mq}$ is directed downwards. Let $\widetilde{Q}_1, \widetilde{Q}_2 \in \mq$ and let 
$A := \{\tau^{\widetilde{Q}_1}_{\rho} \geq  \tau^{\widetilde{Q}_2}_{\rho}\}$, 
$$
\sigma :=  1_A \tau^{\widetilde{Q}_2}_{\rho} + 1_{A^c}T = 1_A \tau^{\widetilde{Q}_1}_{\rho}\wedge\tau^{\widetilde{Q}_2}_{\rho} + 1_{A^c}T, 
$$
and let $\widetilde{Q}_3$ be the pasting of $\widetilde{Q}_1$ and $\widetilde{Q}_2$ in $\sigma$. Then 
$$ 
Z^{\widetilde{Q}_3}_{\tau^{\widetilde{Q}_1}_{\rho}\wedge\tau^{\widetilde{Q}_2}_{\rho}} = Z^{\widetilde{Q}_2}_{\tau^{\widetilde{Q}_2}_{\rho}}\ 1_A + Z^{\widetilde{Q}_1}_{\tau^{\widetilde{Q}_1}_{\rho}} 1_{A^c},
$$
due to Lemma \ref{lccep}. This implies that  $ \tau^{\widetilde{Q}_3}_{\rho} \leq \tau^{\widetilde{Q}_1}_{\rho}\wedge\tau^{\widetilde{Q}_2}_{\rho}$. We conclude the existence of a  sequence
$\{\widetilde{Q}_i\}_{i=1}^{\infty} \subset \mq$ such that 
\eq{eqscatla}
{
\tau^{\widetilde{Q}_i}_{\rho} \searrow \essinf_{Q \in \mq} \tau^Q_{\rho},
}
so that $\tau^{\downarrow}_{\rho}$ is in fact a stopping time.\\ 

\item Let $Q^0 \in \mq$ be arbitrary but fixed. There exists a sequence $\seqa[Q]$  such that $\tau^{{Q}^i}_{\rho} \leq \tau^{\widetilde{Q}^i}_{\rho} \wedge \tau^{Q^0}_{\rho}$ with the further property that
$$
{Q}^i = Q^0 \textrm{ in } \mf_{\tau^{{Q}^i}_{\rho}}.
$$
Indeed, let  $\{\widetilde{Q}_i\}_{i=1}^{\infty}$  be the sequence of probability measures constructed in the previous step. We only need to define  ${Q}^i$ as the pasting of $Q^0$ and $\widetilde{Q}_i$ in the stopping time $\sigma_i$ defined by 
$$
\sigma_i:= 1_{B_i} \tau^{Q^0}_{\rho}\wedge\tau^{\widetilde{Q}_i}_{\rho} +  1_{B^c_i}T,
$$
where $B_i:=\{\tau^{Q^0}_{\rho} \geq \tau^{\widetilde{Q}_i}_{\rho} \}$.

\item  Now we prove (\ref{lespmm}).  Only the inequality 
$$\lsv[\rho] \leq  \essinf_{Q \in \mq}  E_{Q}[H_{\tau^{\downarrow}_{\rho}}\mid \mf_{\rho}],$$
needs a proof. We first note that for any $Q \in \mq$  the inequality $\tau^{\downarrow}_{\rho}\leq \tau^Q_{\rho}$  holds $Q$-a.s. and infer that
\eq{primdes}
{
Z^Q_{\rho} = E_Q[Z^Q_{\tau^{\downarrow}_{\rho}} \mid \mf_{\rho}] \geq E_Q[H_{\tau^{\downarrow}_{\rho}} \mid \mf_{\rho}],
}
where we have used the fact that the random variable $Z^Q_{\tau^{\downarrow}_{\rho}}$ is equal $Q$-a.s. to the Snell envelope of $H$ with respect to $Q$ stopped in $\tau^{\downarrow}_{\rho}$, and the fact that the stopped process $\{U^Q_{\tau^{\downarrow}_{\rho} \wedge s}\}_{s \in [\rho,T]}$ is a $Q$-martingale from time $\rho$ on; see Theorem \ref{theosekaroui}.\\

Recall the sequence  $\seqa[Q]$ constructed in the previous step, so that 
\eq{eqtaupnlm}
{
\tau^{Q^i}_{\rho} \searrow \essinf_{Q \in \mq}  \tau^Q_{\rho} \textrm{ and  } Q^i = Q^0 \textrm{ in } \mf_{\tau^{Q^i}_{\rho}}.
}

By definition of the stopping time $\tau^{Q^i}_{\rho}$, we  have that 
\eq{eqdeveqypbd}
{
Z^{Q^i}_{\tau^{Q^i}_{\rho}} = H_{\tau^{Q^i}_{\rho}}.
}
If we take limits on both sides of  this identity, then we obtain:
\eq{eqdeveqzp}
{
H_{\tau^{\downarrow}_{\rho}} = \lim_{i\rightarrow \infty } H_{\tau^{Q^i}_{\rho}}=\lim_{i\rightarrow \infty} Z^{Q^i}_{\tau^{Q^i}_{\rho}}.
}
In the first equality we have used the fact that the process $H$ is right-continuous, and in the second equality we have used (\ref{eqdeveqypbd}).\\

Now, for $A \in \mf_{\rho}$ the equality (\ref{eqdeveqzp}) develops into
\begin{align} \notag
\int_{A} \lsv[{\rho}] dQ^{0} &\leq \int_{A}\liminf_{i\rightarrow \infty }Z_{\rho}^{Q^{i}}dQ^{0} \notag \\
&\leq \liminf_{i\rightarrow \infty }\int_{A}Z_{\rho}^{Q^{i}}dQ^{0} \label{lfatou}\\
&= \liminf_{i\rightarrow \infty }\int_{A}E_{Q^{i}}[Z_{\tau _{\rho}^{Q^{i}}}^{Q^{i}}\mid \mf_{\rho}]dQ^{0} \label{lidentico}\\ 
&= \liminf_{i\rightarrow \infty }\int_{A}E_{Q^{0}}[Z_{\tau _{\rho}^{Q^{i}}}^{Q^{i}}\mid \mf_{\rho}]dQ^{0} \label{lidenticoo} \\
&=\liminf_{i\rightarrow \infty }\int_{A}Z_{\tau _{\rho}^{Q^{i}}}^{Q^{i}}dQ^{0}  \label{lidenticooo} \\
&=\liminf_{i\rightarrow \infty }\int_{A} H_{\tau _{\rho}^{Q^{i}}}dQ^{0} \label{lporunii}\\
&=\int_{A}H_{\tau^{\downarrow}_{\rho}}dQ^{0} \label{lporuniii} \\
&=\int_{A}E_{Q^{0}}[H_{\tau^{\downarrow}_{\rho}} \mid  \mf_{\rho}]dQ^{0}, \label{lporuniiii}
\end{align}
where the inequality in (\ref{lfatou}) is an application of  Fatou's lemma. The identity in (\ref{lidentico}) follows from the first part of (\ref{primdes}) and (\ref{eqtaupnlm}). The identity (\ref{lidenticoo}) is justified  from the fact that $Q^i=Q^0$ in $\mf_{\tau^{Q^i}_{\rho}}$.
The equality (\ref{lidenticooo}) follows because $A$ is $\mf_{\rho}$-measurable. The equality (\ref{lporunii})  follows from (\ref{eqdeveqypbd}).
In the equality (\ref{lporuniii}) we have applied Lebesgue's convergence theorem, which we are allowed to do justified by (\ref{eqdeveqzp}) and  the fact that  the process $H$ is of  $class(D)$ with respect to $Q^0$. The last equality (\ref{lporuniiii}) follows because $A$ is $\mf_{\rho}$-measurable. Since $Q^0 \in \mq$ was arbitrary we conclude (\ref{lespmm}). 

\item  We still must prove \eqref{tmaxroutst}. This is a consequence of \eqref{lespmm} as we are going to see in Corollary \ref{lmmls} below.\fintheo
\end{enumerate}

The next corollary establishes a  minimax identity. Recall that the lower Snell envelope $\lsv[]$ was defined in Formula \eqref{eqdeflsv}.

\nc{lmmls}
{ 
The following minimax identity
\eq{eqlsvminmax}
{
\lsv[\rho] =\esssup_{\tp \in \ftp[\rho,T]} \essinf_{Q \in \mq} E_Q[H_{\tp} \mid \mf_{\rho}], \quad R-a.s.,
}
holds true. The stopping time $\tau^{\downarrow}_{\rho}$ solves the following robust stopping problem
\eq{eqlsvminmaxrsts}
{
\esssup_{\tp \in \ftp[\rho,T]} \essinf_{Q \in \mq}  E_Q[H_{\tp} \mid \mf_{\rho}].
}
In particular,  for $\rho=0$, $\tau^{\downarrow}_0$ solves the robust stopping problem
\eq{eqminimaxidentitylsvrpi}
{
 \sup_{\tp \in \ftp} \inf_{Q \in \mq}E_Q[H_{\tp}],
}
and 
\eq{eqminimaxidentitylsvrpisolucion}
{
\sup_{\tp \in \ftp} \inf_{Q \in \mq}E_Q[H_{\tp}]
=\inf_{Q \in \mq}\sup_{\tp \in \ftp}E_Q[H_{\tp}].
}
}
Proof.
We show (\ref{eqlsvminmax}). The inequality $\geq$ is obvious. For the converse, note that we have the obvious inequality
$$
\essinf_{Q \in \mq} E_{Q}[H_{\tau^{\downarrow}_{\rho}}\mid \mf_{\rho}]
\leq \esssup_{\tp \in \ftp[\rho,T]} \essinf_{Q \in \mq} E_Q[H_{\tp}\mid \mf_{\rho}],
$$
which together with (\ref{lespmm}) implies that 
$$
\lsv[\rho]= \essinf_{Q \in \mq} E_{Q}[H_{\tau^{\downarrow}_{\rho}}\mid \mf_{\rho}]\leq \esssup_{\tp \in \ftp[\rho,T]} \essinf_{Q \in \mq} E_Q[H_{\tp}\mid \mf_{\rho}] \leq \lsv[\rho].
$$
This establishes (\ref{eqlsvminmax}) and at the same time (\ref{eqlsvminmaxrsts}).\\

The second part of the corollary follows by setting $\rho=0$ in (\ref{eqlsvminmax}) and  (\ref{eqlsvminmaxrsts}).\fincor\\

\nr{remvaluprocess}
{
Note that 
$$
\{ \esssup_{\tp \in \ftp[t,T]} \essinf_{Q \in \mq}  E_Q[H_{\tp} \mid \mf_{t}]\}_{0\leq t \leq T},
$$
is the value process of the robust stopping problem \eqref{eqminimaxidentitylsvrpi}. Corollary \ref{lmmls} implies that this process coincides  with the  lower Snell envelope $\lsv[]$.
}
%************************************************************************************************************************
\section{$\ftp$-Systems} \label{labelsectiontsystems}
%************************************************************************************************************************
In this section we present the concept of $\ftp$-systems and recollect the results we are going to apply for the construction of a right-continous version of the lower Snell envelope.

\nd[$\ftp$-System]{defdellacherie}
{
A family of random variables indexed  by the family of stopping times $\{X(\tp)\}_{\tp \in \ftp}$ is a $\ftp$-{\rm System} if it satisfies the conditions of 
\begin{enumerate}
	\item Adaptedness. For any stopping time $\tp \in \ftp$ the random variable $X(\tp)$ is $\mf_{\tp}$-measurable.
	\item Compatibility.  For any pair of stopping times $\tp^1, \tp^2 \in \ftp$ 
\[
X(\tp^1)=X(\tp^2), R-a.s. \textrm{ in the event } \tp^1=\tp^2.
\]
\end{enumerate}
}
A major topic in \cite{Dellacherielenglartrecollement} is the problem of ``recollement'' of $\ftp$-systems:
\nd{defrecoproc}
{
Let $\{X(\tp)\}_{\tp \in \ftp}$ be a $\ftp$-system.  An optional stochastic process $\h[X]$ {\rm pastes} the $\ftp$-system if for any stopping time $\tp \in \ftp$
\[
X(\tp)=X_{\tp}.
\]
}
Dellacherie and Lenglart considers this problem in greater generality  for \textit{chronologies} $\ftp'\subset \ftp$. They present examples where there is no process pasting a $\ftp'$-system.  However, the next regularity property is sufficient for a $\ftp$-system to be pasted.
\nd{defuppsemcon}
{
A $\ftp$-system $\{X(\tp)\}_{\tp \in \ftp}$ is {\rm upper semicontinuous from the right} if for any decreasing sequence of stopping times $\seqa[\tp]\subset \ftp$ converging to a stopping time $\tp$ we have
\[
X(\tp) \geq \limsup_{i \to \infty} X(\tp^i), R-a.s.
\]
The system is called {\rm lower semicontinuous from the right}, if $\{-X(\tp)\}_{\tp \in \ftp}$ is upper semicontinuous from the right. A system which is both upper and lower semicontinuous from the right is simply said to be {\rm right continuous}.
}
The next theorem solves the problem of ``recollement'' of a $\ftp$-system. It is a difficult  result,  it involves fine results of Bismut and Skalli\cite{Bismutskallitempsarret}, Dellacherie\cite{Dellacherieexistenceessinf}, and   Doob\cite{Doobstopromeas}.
\nt{teodellachrielenglart}
{
Any $\ftp$-system which is upper semicontinuous from the right can be pasted by a unique optional stochastic process whose trajectories are also upper semicontinuous from the right.
}
Proof. See Theorem 4 of Dellacherie and Lenglart.\fintheo\\

The next corollary will allow us to construct a right-continuous version of the lower Snell envelope.
\nc{corrightcontinuity}
{
Any $\ftp$-system which is continuous from the right can be pasted by a unique optional stochastic process whose trajectories are also right continuous.
}
Proof. See the Remark following Corollary 11 of Dellacherie and Lenglart.\fincor\\

%************************************************************************************************************************
\section{Proof of Theorem \ref{teoconstructionse}} \label{labelsectionlaprueba}
%************************************************************************************************************************
\nl{lemmaftpsystem}
{
The family of random variables $\{\lsv[\tp]\}_{\tp \in \ftp}$ is a $\ftp$-system.
}
Proof. The adaptedness of the family is clear due to the definition of the random variable $\lsv[\tp]$. In order to verify the property of  compatibility  we take two stopping times $\tp^1,\tp^2 \in \ftp$. Let us call $A:= \{\tp^1=\tp^2\}$. It is clear that  $A$ is $\mf_{\tp^1\wedge\tp^2}$-measurable. By properties of conditional expectation and essential infimum we have
\[
\lsv[\tp^1]=\lsv[\tp^1] 1_A + \lsv[\tp^1] 1_{A^c}
\]
and
\[
\lsv[\tp^2]=\lsv[\tp^1]1_A + \lsv[\tp^2] 1_{A^c}.
\] 
Thus, $\lsv[\tp^1]=\lsv[\tp^2]$ $R$-a.s. in the event $A$.\finlemma

\nl{lemmaftpsystemrightcontinuous}
{
The $\ftp$-system $\{\lsv[\tp]\}_{\tp \in \ftp}$ is right continuous.
}
Proof. Let $\seqa[\tau] \subset \ftp$ be a decreasing sequence of stopping times converging to $\tau$. We first verify that the $\ftp$-system  is upper semicontinuous from the right.  To this end, let $Q\in \mq$ be fixed but arbitrary. It is clear that
\[
\limsup_{i \to \infty} \lsv[\tau^i] \leq \limsup_{i \to \infty} Z^Q_{\tau^i}.
\]
We have $\limsup_{i \to \infty} Z^Q_{\tau^i} = Z^Q_{\tau}$ due to  the first part of Theorem \ref{theosekaroui}.  Thus,
\[
\limsup_{i \to \infty} \lsv[\tau^i] \leq  Z^Q_{\tau}.
\]
Since $Q$ was arbitrary we conclude that
\[
\limsup_{i \to \infty} \lsv[\tau^i] \leq  \essinf_{Q \in \mq}Z^Q_{\tau}= \lsv[\tau].
\]
This last inequality shows upper semicontinuity.\\

Now we prove lower semicontinuity from the right.\\

In the minimax identity \eqref{eqlsvminmax} of  Corollary \ref{lmmls} we have proved the identity
\[
\lsv[\tau]=\esssup_{\tp \in \ftp[\tau, T]} \essinf_{Q \in \mq}  E_Q[H_{\tp} \mid \mf_{\tau}],
\]
for $\tau \in \ftp$. Then, for  a fixed stopping time $\tp \in \ftp[\tau,T]$ it suffices to establish the inequality 
\eq{eqoppineqsim}
{
\liminf_{i \to \infty} \lsv[\tau^i] \geq \essinf_{Q \in \mq}   E_Q[H_{\tp} \mid \mf_{\tau}].
}
\\

\begin{enumerate}
\item  We prove the inequality
\eq{eqsmplicacionunodostresprim}
{
\lii \essinf_{Q \in \mq}  E_Q[H_{\tp} \mid  \mf_{\tau^i}] \leq \essinf_{Q \in \mq} E_Q[H_{\tp} \mid \mf_{\tau}].
}
For $Q \in \mq$ fixed, we have
\begin{align} \notag
\lii \essinf_{Q \in \mq}   E_Q[H_{\tp} \mid  \mf_{\tau^i}] &\leq \lii   E_Q[H_{\tp} \mid  \mf_{\tau^i}]\\ \notag
&= E_Q[H_{\tp} \mid  \mf_{\tau}],\notag
\end{align}
where the last equality holds true due to Lemma \ref{lemadoobbmar} below, since the  filtration $\mbf$ is right continuous. Thus, if we take the essential infimum over $Q \in \mq$ we obtain \eqref{eqsmplicacionunodostresprim}.

\item  For $Q_0 \in \mq$ arbitrary but fixed, we show
\eq{eqsmplicacionunodostrescuatro}
{
E_{Q_0}[\lii \essinf_{Q \in \mq}  E_Q[H_{\tp} \mid  \mf_{\tau^i}]] \geq E_{Q_0}[\essinf_{Q \in \mq}  E_Q[H_{\tp} \mid \mf_{\tau}]],
}

The  sequence of random variables  
\eq{eqbacksubmartingaleproof}
{
\seq[Y]:=\{ \essinf_{Q \in \mq}  E_Q[H_{\tp} \mid  \mf_{\tau^i}]\}_{i=1}^{\infty}
}
is a Backwards-submartingale for each $Q \in \mq$, due to  Lemma  \ref{lemabackwardssubmarrobust}. This same result yields that the limit inferior in \eqref{eqsmplicacionunodostresprim} actually exists as a limit.  Then, we get:
\begin{align} \notag
E_{Q_0}[ \lii \essinf_{Q \in \mq}   &E_Q[H_{\tp} \mid  \mf_{\tau^i}]] \\
=& E_{Q_0}[ \lis \essinf_{Q \in \mq}   E_Q[H_{\tp} \mid  \mf_{\tau^i}]] \label{primeecuacionenlapartefinalunodos} \\
\geq& \lis E_{Q_0}[ \essinf_{Q \in \mq}   E_Q[H_{\tp} \mid  \mf_{\tau^i}]]. \label{primeecuacionenlapartefinalunotres}
\end{align}
In (\ref{primeecuacionenlapartefinalunodos}) we have used the fact that the limit exists. In (\ref{primeecuacionenlapartefinalunotres}) we have used Fatou's lemma, which we are allowed to apply  since the sequence $\seq[Y]$  is,  obviously,  uniformly integrable with respect to $Q_0$. To conclude \eqref{eqsmplicacionunodostrescuatro} we show
\eq{primeecuacionenlapartefinalunocuatro}
{
\lis E_{Q_0}[ \essinf_{Q \in \mq}   E_Q[H_{\tp} \mid  \mf_{\tau^i}]]
\geq 
E_{Q_0}[\essinf_{Q \in \mq}  E_Q[H_{\tp} \mid \mf_{\tau}]].
}
For a stopping time $s \in \ftp$,  recall the notation
$$
\mq(Q_0,s) = \{Q \in \mq \mid Q= Q_0 \textrm{ in } \mf_{s}\}.$$
We  observe that 
\[
E_{Q_0}[ \essinf_{Q \in \mq}   E_Q[H_{\tp} \mid  \mf_{\tau^i}]]= \inf_{Q \in \mq(Q_0,\tau^i)}E_Q[H_{\tp}]
\]
and 
\[
E_{Q_0}[ \essinf_{Q \in \mq}   E_Q[H_{\tp} \mid  \mf_{\tau}]]= \inf_{Q \in \mq(Q_0,\tau)}E_Q[H_{\tp}],
\]
due to Lemma \ref{lfplll}. Note that 
$$\mq(Q_0,\tau^i) \subset \mq(Q_0,\tau).$$
Let $\epsilon >0$ and let $Q^i \in \mq(Q_0,\tau^i)$ be such that 
\[
E_{Q^i}[H_{\tp}]- \epsilon  \leq \inf_{Q \in \mq(Q_0,\tau^i)}E_Q[H_{\tp}].
\]
The inequality \eqref{primeecuacionenlapartefinalunocuatro} will follow from 
\eq{equltimareducciontrivial}
{
\lis E_{Q^i}[H_{\tp}] \geq \inf_{Q \in \mq(Q_0,\tau)}E_Q[H_{\tp}],
}
but  $Q^i \in  \mq(Q_0,\tau)$ so that 
\[
E_{Q^i}[H_{\tp}] \geq \inf_{Q \in \mq(Q_0,\tau)}E_Q[H_{\tp}],
\]
implying (\ref{equltimareducciontrivial}).

\item The inequalities \eqref{eqsmplicacionunodostresprim} and \eqref{eqsmplicacionunodostrescuatro} imply the identity
\eq{eqsmplicacionunodostresprimtrescuatro}
{
\lii \essinf_{Q \in \mq}  E_Q[H_{\tp} \mid  \mf_{\tau^i}] = \essinf_{Q \in \mq} E_Q[H_{\tp} \mid \mf_{\tau}].
}

\item In this step we reduce the proof of  \eqref{eqoppineqsim} to  \eqref{eqsmplicacionunodostresprimtrescuatro}. We define 
\[
\tp^{(i)}:= \tp \indf[{\tp \geq \tau^i}] + T \indf[{\tp < \tau^i}] \in \ftp[\tau^i,T].
\]
Then we get 
\[
\lsv[\tau^i] \geq \essinf_{Q \in \mq} E_Q[H_{\tp^{(i)}} \mid \mf_{\tau^i}],
\]
so that
\[
\lii \lsv[\tau^i] \geq \lii \essinf_{Q \in \mq}  E_Q[H_{\tp^{(i)}} \mid \mf_{\tau^i}].
\]
To prove (\ref{eqoppineqsim}) it is enough to show that 
\eq{eqsmplicacionuno}
{
\lii \essinf_{Q \in \mq}  E_Q[H_{\tp^{(i)}} \mid \mf_{\tau^i}] \geq \essinf_{Q \in \mq} E_Q[H_{\tp} \mid \mf_{\tau}].
}
We simplify the proof of (\ref{eqsmplicacionuno}). Note that  
\[
E_Q[H_{\tp^{(i)}} \mid  \mf_{\tau^i}]=  1_{\{\tp \geq \tau^i\}}E_Q[H_{\tp} \mid  \mf_{\tau^i}] + 1_{\{\tp < \tau^i\}}E_Q[H_T \mid  \mf_{\tau^i}],
\]
so that (\ref{eqsmplicacionuno}) will follow from  the next inequality
\eq{eqsmplicacionunodos}
{
\lii \essinf_{Q \in \mq} 1_{\{\tp \geq \tau^i\}}E_Q[H_{\tp} \mid  \mf_{\tau^i}] \geq \essinf_{Q \in \mq} E_Q[H_{\tp} \mid \mf_{\tau}].
}
Since $R(\li1_{\{\tp \geq \tau^i\}} =1)=1$ monotonously, then we can simplify the proof of  (\ref{eqsmplicacionunodos}) into the proof of the following inequality
$$
\lii \essinf_{Q \in \mq} E_Q[H_{\tp} \mid  \mf_{\tau^i}] \geq \essinf_{Q \in \mq} E_Q[H_{\tp} \mid \mf_{\tau}],
$$
which we know holds true due to \eqref{eqsmplicacionunodostresprimtrescuatro}.\finlemma
\end{enumerate}

\nl{lemadoobbmar}
{
Let  $Y$ be a positive  random variable such that
\[
E_R[Y] < \infty.
\]
Let $\seq[\mf]$ be a decreasing sequence  of sub-$\sigma$-algebras of $\mf$, that is, $\mf_{i+1} \subset \mf_{i} \subset \mf$. Then 
\[
\li E_R[Y \mid \mf_i ] = E_R[Y \mid \mf_{-\infty}],
\]
where $\mf_{-\infty} = \cap_{i=1}^{\infty} \mf_i$.
}
Proof.
This is a special case of the Backwards-martingale convergence theorem; see e.g., Theorem 2.I.5, or Theorem 2.III.16  in Doob\cite{Doobpotentialtheory}.\finlemma\\

Now we conclude the proof of Theorem \ref{teoconstructionse} as follows.\\

Proof. The family of random variables 
\[
\{\lsv[\tp]\}_{\tp \in \ftp}
\]
is a $\ftp$-system, due to Lemma \ref{lemmaftpsystem}. Moreover, this system is  right continuous, due to Lemma \ref{lemmaftpsystemrightcontinuous}. The theorem now follows from Corollary \ref{corrightcontinuity}.\fintheo\\

%------------------------------------------------------------------------------------------------------------------------

%------------------------------------------------------------------------------------------------------------------------
\end{document}